\documentclass[aps,prb,superscriptaddress]{revtex4-2}

\usepackage{parskip}
\usepackage[utf8]{inputenc}
\usepackage[T1]{fontenc}
\usepackage[english]{babel}
\usepackage{hhline}
\usepackage{amsmath}
\usepackage{wasysym}
\usepackage{amsfonts}
\usepackage{amssymb}
\usepackage{graphicx}
\usepackage{float}
\usepackage{url}
\usepackage{nicefrac}
\usepackage{braket}
\usepackage{amsmath}
\usepackage{slashed}
\usepackage{blkarray}
\usepackage{multirow}
\usepackage{hyperref}
\usepackage{color}
\numberwithin{equation}{section}

\begin{document}
\title{Electric-field gradients at the nuclei from all-electron, four-component relativistic density-functional theory using Gaussian-type orbitals}
\date{\today}
\author{Marc Joosten}
\email{marc.joosten@uit.no}
\affiliation{Hylleraas Center for Quantum Molecular Sciences, Department of Chemistry, University of Troms\o ---The Arctic University of Norway, 9037 Troms\o , Norway}
\author{Michal Repisky}
\email{michal.repisky@uit.no}
\affiliation{Hylleraas Center for Quantum Molecular Sciences, Department of Chemistry, University of Troms\o ---The Arctic University of Norway, 9037 Troms\o , Norway}
\affiliation{Department of Physical and Theoretical Chemistry, Faculty of Natural Sciences, Comenius University, Ilkovicova 6, SK-84215 Bratislava, Slovakia}
\author{Marius Kadek}
\email{marius.kadek@uit.no}
\affiliation{Hylleraas Center for Quantum Molecular Sciences, Department of Chemistry, University of Troms\o ---The Arctic University of Norway, 9037 Troms\o , Norway}
\author{Pekka Pyykk\"{o}}
\email{pekka.pyykko@helsinki.fi}
\affiliation{Department of Chemistry, P.O. Box 55, FIN-00014, University of Helsinki,
Finland}
\author{Kenneth Ruud}
\email{kenneth.ruud@uit.no}
\affiliation{Hylleraas Center for Quantum Molecular Sciences, Department of Chemistry, University of Troms\o ---The Arctic University of Norway, 9037 Troms\o , Norway}
\affiliation{Norwegian Defence Research Establishment, Instituttveien 20, 2027 Kjeller, Norway}

\begin{abstract}
  We present an all-electron, four-component relativistic implementation of
  electric field gradients (EFGs) at the nuclei using Gaussian-type
  orbitals and periodic boundary conditions. This allows us to include
  relativistic effects variationally, which is important for compounds
  containing heavy elements and for a property dependent the electronic
  structure close to the nuclei. The all-electron approach ensures 
  an accurate treatment of both  core and valence orbitals, as both are
  important in the evaluation of EFGs. Computational efficiency is achieved
  through the use of a recent implementation of density fitting in
  combination with quaternion algebra and restricted kinetic balance. We use the relativistic approach to calculate
  the EFGs in different arsenic, antimony and bismuth halides and oxyhalides,
  and explore the importance of relativistic effects on EFGs in solids and compare these with results obtained for molecular species.  Our
  calculations contribute to establishing a reliable estimate for the nuclear quadrupole moment of
  $^{209}\mathrm{Bi}$, for which our best estimate is $-428(17)$ mb, in
  excellent agreement both with molecular data and a recent reevaluation of the
  nuclear quadrupole moment obtained from atomic data and {\it ab initio\/} calculations.  Our
  results suggest that there is a need to revisit the experimental data for the
  EFGs of several bismuth oxyhalides.
\end{abstract}
\maketitle

\section{Introduction}
\label{sec:introduction}

Nuclear electric quadrupole moments (NQMs) are important parameters providing
fundamental information about nuclear structure, in particular whether the
nuclear structure is spherically symmetric or not. A prerequisite for the
existence of a non-vanishing NQM is a nuclear spin larger than $1/2$. The
nuclear quadrupole moments interact with the electric field gradients (EFGs) at
the nuclei, leading to a splitting of atomic or molecular energy levels that
can be observed for instance in nuclear quadrupole resonance
spectroscopy~\cite{smith1971nuclear} or rotational
spectroscopy~\cite{puzzarini2010quantum,puzzarini2013rotational} and correspond
to excitation energies in the microwave region.
Through measurements of the coupling of NQMs and EFGs, one can also
obtain information about the electron density close to the nuclei, provided
that accurate values of the NQMs are available. Much effort has therefore
been devoted to determine NQMs~\cite{pyykko2018year}.

One way of determining NQMs is to combine accurate {\em ab initio\/}
calculations of EFGs with highly accurate transition energy data obtained from
different experiments, and derive the NQMs using the relation between the EFGs
and transition energies. A number of NQMs have been determined with very high
accuracy using this hybrid experimental/computational approach.  For further
information on various methods of determining NQMs, their expected accuracy,
and the most up-to-date list of NQMs, we recommend the review by
Pyykk\"{o}~\cite{pyykko2018year}.

EFGs probe the electron density close to the nuclei, making relativistic
effects crucial, particularly for heavier nuclei~\cite{van2000density}. The
calculation of EFGs is relatively straightforward, since the EFGs are
first-order molecular properties. They can be evaluated either as expectation
values for variational wave functions or by employing  Z-matrix or Lagrangian techniques for
highly correlated, non-variational wave functions~\cite{handy1984on,helgaker1989configuration}, such as 
coupled-cluster or configuration interaction methods. 
The incorporation of relativistic effects is also fairly straightforward, even at the fully
relativistic level of theory using highly correlated wave
functions~\cite{visscher1998molecular,de1998bonding}. However, when relativistic
effects are approximated using a transformed Hamiltonian, significant
picture-change effects must be accounted for~\cite{autschbach2012two}. We also note that
special care must be taken if a finite-field approach is applied as an
alternative way of calculating  EFGs.~\cite{pernpointner1998point}

Compounds containing heavy elements often exist as solids. In the field of
computational materials modeling, a plane-wave description of the valence
electrons is often combined with the pseudopotential technique where the
oscillatory core electronic orbitals are effectively substituted with
pseudo-wave-functions that mimic the effect of the core electrons on the
valence electrons without the intricate oscillatory pattern.  The use of
pseudopotentials allows relativistic effects on the core electrons to easily be
included, but the approximate description of the core electron density makes
them unsuited for evaluating properties such as the EFGs. The projector
augmented wave method allows for calculations of  EFGs at the
nuclei~\cite{petrilli1998electric,profeta2003accurate}, but reaching accurate
results can require that the EFGs are corrected  using molecular calculations
at a more advanced theory level (\emph{e.g.} with a hybrid exchange-correlation
functional)~\cite{hartman2021fast}. Alternative relativistic approaches include
the use of four-component
numerical~\cite{jeglivc2009influence,koch2009analysis,koch2009crystal} or
two-component Slater-type atomic
orbitals~\cite{van2000density,autschbach2010analysis} or techniques that
combine plane waves with auxiliary functions for explicit treatment of core
electrons, while potentially also including relativistic
effects~\cite{blaha1985first,petrilli1998electric,pyykko2018year}.
Nevertheless, such hybrid approaches are inherently implementationally
complex and computationally intensive, necessitating meticulous convergence
settings and a profound understanding of the methods to ensure the accuracy of the calculated results
in complex systems.

We have adopted an alternative, efficient all-electron approach that
incorporates relativistic effects variationally at the four-component level of
theory and which utilizes Gaussian-type basis sets with periodic boundary conditions,
enabling systematic improvements in the results as the basis set is expanded.
Computational efficiency is ensured by the use of quaternion algebra and
restricted kinetic balance~\cite{repisky2020respect,kadek2019all}, as well as
the resolution-of-the-identity approximation~\cite{eichkorn1995auxiliary,eichkorn1997auxiliary} for the Coulomb integrals in
combination with multipole expansion of the far-field
contributions~\cite{JoostenRIJ}.  Our previous work has shown that our method
reliably achieves accurate band gaps already with double-zeta quality basis
sets, and that smooth convergence of the density is assured without issues even
when diffuse functions are incorporated into the basis set~\cite{kadek2019all}.
Indeed, we have established that retaining diffuse functions is essential for
rapid basis set convergence; conversely, the more commonly employed ``reduced''
basis sets (where the diffuse functions are removed)~\cite{peralta2005scalar,perry2001antiferromagnetic,usvyat2011approaching} yield inferior results and
slower basis set convergence.

We note here that the use of Gaussian-type orbitals (GTOs) for
computational studies of periodic solids has an extensive history at the
non-relativistic level of theory, pioneered by the work of Pisani and
Dovesi~\cite{pisani1980exact,dovesi1983treatment}, followed by several more
recent
implementations~\cite{lippert1997hybrid,boettger2000inclusion,kudin2000linear,
lazarski2015density,becker2019density,sun2017gaussian}.  Some of these
GTO-based codes also reported calculations of EFGs for solid-state
systems~\cite{PalmerBlairFish+1994+137+145,stare2020nuclear,
declerck2006first,schmidt2008beyond,sun2020recent}, albeit without the full
consideration of relativistic effects.

Here, we extend our relativistic all-electron approach based on GTOs to the
calculation of EFGs in solids and use the first-principles calculations in
combination with experimental data to determine the NQM of As, Sb, and Bi from
the EFGs in their halides and oxyhalides. Furthermore, we investigate the
discrepancy between the atomic and molecular values of the NQM of $^{209}$Bi by
using solid-state Nuclear Quadrupole Resonance (NQR) data.

The remainder of the paper is organized as follows: In Section~\ref{sec:theory},
we provide the basic theory for the calculation of EFGs in solids at the
four-component level of theory using Gaussian-type orbitals. In
Section~\ref{sec:compdetails}, we give the computational details for the
calculations presented in this work. In Section~\ref{sec:results}, we explore
the importance of relativistic effects and choice of basis sets on the EFGs in a
series of pnicogen halides and oxyhalides, with a particular focus on
bismuth-containing systems. We obtain an estimate of the NQM of
$^{209}$Bi and compare it with atomic and molecular data. Finally, in
Section~\ref{sec:conclusions}, we give some concluding remarks and an outlook.

\section{Theory}
\label{sec:theory}

NQR utilizes the interaction of the
nuclear charge density with the surrounding electric field. The quadrupole
interaction is the leading contribution, \emph{i.e.} the interaction between
the nuclear quadrupole moment and the electric field gradient tensor.  A
detailed discussion can be found in Ref.~\cite{seliger2000nuclear}. This leads to the NQR Hamiltonian in the eigensystem of the EFG, given by
\begin{equation}
H_{NQR} = \frac{Q}{2I(2I-1)}\left(V^{xx}I^{x} + V^{yy}I^{y} + V^{zz}I^{z}\right) =\frac{qQ}{4I(2I-1)}\left(3I^z-I^2 + \frac{\eta}{2} \left(I^2_+ + I^2_-\right)\right).
\end{equation}
We have here introduced the eigenvalues of the electric field gradient components defined according to $|V^{zz}| > |V^{yy}| > |V^{xx}|$,
the asymmetry parameter $\eta = (V^{xx}-V^{yy})/V^{zz}$ and the nuclear spin
$I$. $Q$ denotes the nuclear quadrupole moment (NQM) and $q$ is the largest
eigenvalue $V^{zz}$ and is often referred to as just the EFG.  Diagonalization
then leads to different spin energy levels, with the magnitude depending on $q$
and $Q$, while the general structure depends on $\eta$ and the total spin of
the nucleus $I$. In the following, only the traceless part of the EFG is
calculated; the traced part leads to a constant shift in the spectra
\cite{Thyssen} that is not considered in this work.

The EFG tensor $V^{ij}$ is defined as the second derivative of the electric
potential at the position of the reference nucleus $K$ (using the notation $\partial^i\equiv
\partial/\partial r^i$):
\begin{equation}\label{defV}
V_K^{ij} = \left.\partial^i \partial^j \int d\bold{r'} \rho(\bold{r'}) \frac{1}{\lvert\bold{r'} - \bold{r}\rvert}\right|_{\bold{r}=\bold{r}_K }=\int d\bold{r'} \rho(\bold{r'}) v^{ij}\left(\bold{r'}-\bold{r}_K\right) ,
\end{equation}
with
\begin{equation}
v^{ij}(\bold{r}) = \frac{3 r^i r^j-\delta^{ij} r^2}{r^5},
\end{equation} 
and the total charge density of the infinite periodic system 
\begin{equation}
\rho(\bold{r}) = \sum_\bold{m} \rho_\bold{m}(\bold{r})=\sum_\bold{m} \rho_\bold{0}\left(\bold{r}-\bold{r}_\bold{m}\right),
\end{equation}
where $\bold{m}$ labels the unit cell centered at $\bold{r}_\bold{m}$.  In
applying the derivative in Eq.~\eqref{defV}, the Poisson term is neglected.
This has no effect on the traceless part of $V$ \cite{Thyssen}.
Equation \eqref{defV} can be evaluated as stated by keeping the nucleus $K$
fixed and summing over the charge densities of every unit cell. Alternatively,
the summation can be shifted to the nucleus by using translation symmetry
\begin{equation}\label{fullcharge}
V_K^{ij} = \int d\bold{r'} \rho_\bold{0}(\bold{r'}) \sum_\bold{m} v^{ij}\left(\bold{r'}-\bold{r}_{K}+\bold{r}_\bold{m}\right) = \sum_\bold{m} \int d\bold{r'} \rho_\bold{0}(\bold{r'}) v_{\bold{m},K}^{ij}\left(\bold{r'}\right).
\end{equation}
The EFG consists of contributions from the electronic charge density and from all other non-reference nuclei.\\
Expressing the electron
density using the one-electron reduced density matrix $D$ and Gaussian-type
basis functions $\chi_{\mu\bold{n}}$ centered on each unit cell $\bold{n}$, the electronic part of the EFG tensor is obtained as
\begin{equation}\label{elec}
V_{e,K}^{ij}= -\sum_\bold{m} \sum_{\mu \nu \bold{n}} \int d\bold{r} \text{Tr}\left[\chi^\dagger_{\mu \bold{0}}(\bold{r}) v_{\bold{m},K}^{ij}\left(\bold{r} \right)\chi_{\nu \bold{n}}(\bold{r}) D^{\nu\bold{n},\mu \bold{0}}\right] = -\sum_{\bold{m}u} \text{Tr}\left[ I_{\bold{m},K,u}^{ij} D^{\bar{u}} \right],
\end{equation}
where $u\equiv(\mu\nu\bold{n})$, $\bar{u}\equiv(\nu\bold{n}\mu)$, and the EFG
integrals are labeled as $I$. Note here that the basis functions
$\chi_{\nu\bold{n}}$ are four-component Dirac bispinors, and the trace Tr needs
to be introduced to sum over the four components, \emph{i.e.} four diagonal
matrix elements in the product of $I$ and $D$. More details about the
implementation of the calculation of the relativistic electronic structure of solids in the ReSpect program is
found in Ref. \cite{kadek2019all}. Similarly, the nuclear part is given by
\begin{equation}\label{nucl}
V_{n,K}^{ij} = \sum_\bold{m} \sum_{A} Z_A \frac{3\left(\bold{r}_A-\bold{r}_{\bold{m},K}\right)^i\left(\bold{r}_A-\bold{r}_{\bold{m},K}\right)^j - \left|\bold{r}_A-\bold{r}_{\bold{m},K}\right|^2 \delta^{ij} }{\left|\bold{r}_A-\bold{r}_{\bold{m},K}\right|^5}.
\end{equation}
In the summations over all atoms $A$ and unit cells $\bold{m}$, the reference
nucleus term $A=K$ must be excluded for $\bold{m}=\bold{0}$.

For sufficiently large values of $\bold{m}$, the charge density of the unit
cell $\bold{m}=\bold{0}$ has no significant overlap with the reference nuclei
in the cell $\bold{m}$ \footnote{Due to the use of GTOs, the charge density
decays exponentially.}.  Therefore the lattice sum in Eq.~\eqref{fullcharge}
can be split into  near and far-field (NF and FF) parts. The NF is calculated
directly using Eqs.~\eqref{elec} and~\eqref{nucl}. The FF terms can be expanded
using a multipole expansion. This is based on
\begin{multline}\label{mmexp}
\partial^i\partial^j \frac{1}{\left|\bold{r'} - \bold{r}_K - \bold{r}_\bold{m}\right|}=\partial^i\partial^j \frac{1}{\left|\left(\bold{r'} - \bold{r}_C\right) - \left( \bold{r}_K + \bold{r}_\bold{m} - \bold{r}_C\right)\right|} \\ = \sum_{lm} \partial^i\partial^jR_{lm}\left(\bold{r'} - \bold{r}_C\right) I^*_{lm}\left( \bold{r}_K + \bold{r}_\bold{m} - \bold{r}_C\right) =\sum_{lm} R_{lm}\left(\bold{r'} - \bold{r}_C\right) \partial^i\partial^jI^*_{lm}\left( \bold{r}_K + \bold{r}_\bold{m} - \bold{r}_C\right).
\end{multline}
The center of the unit cell is chosen as the center of the multipole expansion
$\bold{r}_C$ and $R$ and $I$ refer to the scaled regular and irregular solid
harmonics as defined in Ref. \cite{helgaker2014molecular}. The derivative in~\eqref{mmexp} acting on $\bold{r'}$ is equivalent to a derivative on $\bold{r}_\bold{m}$ up to a minus sign. This can be used to shift the derivatives between $R$ and $I$. Recursive expressions for
the derivatives are defined in Ref.~\cite{perez1996concise}, albeit using a different sign convention.
The solid harmonics defined above are complex functions. However, in practical
calculations, it is more convenient to use real
harmonics~\cite{helgaker2014molecular}.  Thus, the FF contribution to the EFG
becomes
\begin{equation}\label{VFF}
V^{ij}_{K,FF} = \sum_{lm} 2 \frac{1}{2^{\delta_{m0}}}	Q_{lm}\left(\bold{r}_C\right) \partial^i \partial^j\sum_{\bold{m} \in FF} I_{lm}\left( \bold{r}_K + \bold{r}_\bold{m} - \bold{r}_C\right)=\sum_{lm} 	Q_{lm}\left(\bold{r}_C\right) I^{ij}_{K,FF,lm}\left(\bold{r}_C\right),
\end{equation}
with the total multipole moments of the unit cell defined as
\begin{equation}
Q_{lm}\left(\bold{r}_C\right) = \int d\bold{r}\rho_\bold{0}(\bold{r})R_{lm}\left(\bold{r} - \bold{r}_C\right).
\end{equation}
Eq.~\eqref{VFF} contains both electronic and nuclear contributions and is
evaluated in a similar manner as the NF terms in Eqs.~\eqref{elec}
and~\eqref{nucl}. The additional factor compensates for the switch to real
harmonics and is absorbed into $I$.  The multipole moments are independent both
of the lattice vector $\bold{m}$ and the specific nucleus $K$ and only need to
be calculated once for a given system.  To evaluate the lattice sum, the
renormalization scheme from Ref. \cite{kudin2004revisiting} is used, with
additional details provided in Ref. \cite{kadek2019all}. This scheme only needs to be
employed once per system as it can be shifted to the position of a different
nucleus $K'$ using the translation properties of $I$. Hence, the computational cost
of evaluating the FF terms is negligible, and the entire computational effort
comes from the direct calculation of the NF contributions in Eq.~\eqref{elec}.

The calculation of EFGs for molecules is significantly simpler. Eqs.~\eqref{elec}
and~\eqref{nucl} can then be used directly by removing the lattice indices and the summation over $\bold{m}$. The NF-FF split is in this case  not needed. Our implementation in ReSpect agrees for four-component molecular systems with DIRAC\cite{DIRAC24} and for one-component solids with CRYSTAL\cite{dovesi2020crystal}.


\section{Computational details}
\label{sec:compdetails}
We use the four-component Dirac--Coulomb Kohn--Sham level of theory together
with the generalized gradient approximation (GGA) for the nonrelativistic
exchange--correlation functional (PBE)~\cite{perdew1996generalized}. The fully
uncontracted triple-$\zeta$ (TZ) Dyall basis sets are
used unless otherwise noted~\cite{dyall2002relativistic,dyall2006relativistic}. To accelerate the calculation of the Coulomb electrostatic
interactions, we employed the resolution of the identity
approach~\cite{eichkorn1995auxiliary,eichkorn1997auxiliary,JoostenRIJ}. A detailed discussion of the implementation and accuracy of the RI-J
approximation used in this work will be provided elsewhere~\cite{JoostenRIJ},
but calculations on BiOCl lead to errors of the order of 10$^{-3}$ V/m$^2$ in the EFG
both at the non-relativistic and relativistic levels of theory, a relative
error that is less than 0.01\%. 

The momentum space was sampled with a $9\times
9\times 9$ mesh of $\bold{k}$ points for the bismuth oxyhalides and a $5\times
5\times 5$ mesh for the other systems. The convergence with respect to the
$\bold{k}$-point mesh was assessed by comparing the EFG results obtained from
calculations across a progressively increasing sequence of meshes.

\section{Results and discussion}
\label{sec:results}

In order to demonstrate the importance of relativistic effects for EFGs, we
will have a particular focus on $^{209}$Bi, as there until recently was a
rather significant discrepancy between the NQMs established based on comparing
experimental and computed data between molecular \cite{teodoro2013nuclear} and atomic
studies~\cite{bieron2001nuclear}. This discrepancy was however recently resolved~\cite{dognon2023determining,eliav2023},
bringing the atomic NQMs into very good agreement with the molecular NQMs. We also note that nuclear calculations by Karayonchev {\it et al.\/} support the molecular NQM, with a NQM of $-428$ mbarn~\cite{karayonchev2019lifetimes}. It
would be of interest to see whether our solid-state implementation would
confirm these predictions.

In order to better assess the accuracy of our calculated results, we have performed benchmark calculations on  molecular BiCl$_3$ and solid BiOCl using Dyall's double (DZ), triple (TZ )and quadrupole-$\zeta$ (QZ) basis sets.~\cite{dyall2002relativistic,dyall2006relativistic}. We have also included truncated versions of these basis sets for the solid, in which we apply the commonly used approximation of removing all basis functions with exponents
$<0.1$ when calculating properties of periodic systems using GTOs~\cite{peralta2005scalar,usvyat2011approaching,perry2001antiferromagnetic}.

\begin{table}
\caption{Basis set convergence of Bi EFGs. Results for basis $X$ reported as EFG($X$)/EFG(QZ). For details, see text.}
  \label{tab:basis}
\begin{tabular}{lrrrrrr}\hline\hline
System      & tDz & DZ & tTz & TZ & tQZ & QZ \\ \hline
BiCl$_3$    &- &0.979 &- & 0.997&- & 1   \\
BiOCl       & 0.879&0.938 & 0.984&0.986 & 1.0003& 1  \\\hline\hline
\end{tabular}
\end{table} 

The value of the Bi EFG for the different basis sets are reported in Table~\ref{tab:basis} relative to the Bi EFG obtained using the QZ basis set. BiCl$_3$ uses the isolated molecular structure ({\it vide infra\/}), whereas for BiOCl we use the experimental solid-state structure of Ref.  \citep{BiOX}.  

For the molecular system, both the DZ and TZ results are very close to the QZ value, being 97.9\% and 99.7\% of the QZ value, respectively. For the solid, convergence is worse,  with the  DZ results clearly being of inferior quality, with only 93.8\% of the QZ value, whereas TZ is performing reasonably well with 98.6\% of the QZ value. The Dyall basis sets are optimized for molecular calculations, and a poorer basis set convergence in solids is not surprising.
The truncated DZ basis set performed poorly, whereas for the TZ basis, the truncated result is only 0.2\% off the full basis, and they are identical at the QZ level.

We recently demonstrated that the removal of diffuse functions is not necessary and inhibits basis-set convergence~\cite{kadek2019all,kadek2023band} and prevents convergence to the basis-set limit altogether for band gaps and structures.
Even though EFGs are a core-electron property, this is not the reason for the reduced basis sets not affecting the quality of the calculated results. Instead, only two diffuse basis functions are removed on Bi, whereas in our previous work, as many as 6 basis functions were removed in the case of tungsten.
This highlights that truncation is highly dependent on the systems studied and the basis sets used, and should in general be applied with care and ideally not at all.
In this work, all exponents of the original basis set are retained.

For BiOCl, we have also
performed test calculations with the BLYP functional~\cite{becke1988density,lee1988development,miehlich1989results}. The results are in good agreement with those obtained using the PBE functional, being less
than 3\%\ larger than the PBE results both at the nonrelativistic and
relativistic levels of theory for the  EFG at the Bi nucleus using the
TZ basis set.

Considering the possible uncertainty in the NQM of the $^{209}$Bi nucleus, we
have included a few arsenic and antimony halides and oxyhalides for which there
exist experimental data, in order to benchmark our results against reliable
experimental data. This also allows us to explore the relative importance of
relativistic corrections as we go down in the periodic table. All our results
are collected in Table~\ref{tab:results}. We report, in addition to the
calculated EFG value and the corresponding value for the asymmetry parameter
$\eta$, also the derived value for the NQM using the experimental resonance
frequencies. The reference values for the $^{75}$As, $^{121}$Sb, $^{123}$Sb and $^{209}$Bi
NQMs are -311, -543, -692 and -420 mbarn, respectively~\cite{demovic2010quadrupole,haiduke2006nuclear,barzakh2021large}, and we also  report the ratio
between the calculated and reference values for the NQM.

\begin{table}
  \caption{Calculated EFGs [$10^{21}$ V/$m^2$], asymmetry parameters $\eta$ and resulting $^{209}$Bi quadrupole moments Q [mb].}
  \label{tab:results}
  \begin{tabular}{lcrrrrrrrrrc}\hline\hline
    Compound && \multicolumn{4}{c}{Non-relativistic} &
    \multicolumn{4}{c}{Relativistic} & Exp. & Ref.\\
     & geom & EFG & $\eta$ & Q & Q$_{\mathrm{calc}}$/Q$_{\mathrm{ref}}$ & EFG
    & $\eta$ & Q & Q$_{\mathrm{calc}}$/Q$_{\mathrm{ref}}$ & $\eta_{\mathrm{exp}}$\\\hline
    AsCl$_3$&\cite{AsClgeom} & -19.11 & 0.0 & -341.7 & 1.10 & -18.47 & 0.0 & -353.5 &
    1.14 & 0\footnote{Assumed to be zero.} &\cite{Asdata}\\
    AsBr$_3$&\cite{AsBrgeom} & -14.47 & 0.076 & -349.7 & 1.12 & -13.65 & 0.082 &
    -370.7 & 1.19 & 0$^{\mathrm{a}}$ &\cite{Asdata}\\\\
    $^{121}$SbCl$_3$&\cite{SbClgeom} & -26.87 & 0.151 & -590.6 & 1.09 & -25.60 & 0.170 &
    -619.9 & 1.14 & 0.184&\cite{SbCldata}\\
    $\alpha$-$^{121}$SbBr$_3$&\cite{aSbBrgeom} & -22.63 & 0.114 & -604.9 & 1.08 & -20.76 & 0.155 &
    -659.3 & 1.21 & 0.08&\cite{SbBrdata}\\
    $\beta$-$^{121}$SbBr$_3$\footnote{Exp data only available at 304K}&\cite{bSbBrgeom} & -23.31 & 0.178 & -560.7 & 1.03 & -21.50 & 0.202 &
    -607.9 & 1.12 & 0.18&\cite{SbBrdata}\\
    $^{123}$Sb$_4$O$_5$Cl$_2$ &\cite{SbOClgeom}& -32.81 & 0.261 & -805.0 & 1.16 & -33.37 &
    0.302 & -791.5 & 1.14 & 0.307&\cite{BiOX}\\
    && -34.77 & 0.380 & -805.0 & 1.16 & -35.40 & 0.411 & -791.4 & 1.14
    & 0.396&\cite{BiOX}\\\\
    BiCl$_3$ &\cite{BiClgeom}& -31.95 & 0.362 & -421.3 & 1.00 & -27.36 & 0.565 &
    -492.1 & 1.17 & 0.583&\cite{BiCldata}\\
    BiBr$_3$ &\cite{BiBrgeom}& -26.90 & 0.446 & -523.5 & 1.25 & -21.81 & 0.850 &
    -645.8 & 1.54 & 0.553&\cite{BiBrdata}\\
    BiBr$_3$\footnote{Refitted experimental data, see text.} &\cite{BiBrgeom}&  &  & -387.1
 & 0.92
 & 
 &  & -477.6
     & 1.14 & 0.840&\\\\
    BiOF &\cite{BiOX} & -29.90 & 0.000 & -410.1 & 0.98 & -26.70 & 0.000 & -459.3 &
    1.09 & 0.0&\cite{BiOX}\\
    BiOCl &\cite{BiOX}& -15.39 & 0.000 & -424.8 & 1.01 & -10.62 & 0.001 & -615.8 &
    1.47 & 0.0&\cite{BiOX}\\
    BiOBr &\cite{BiOX}& -12.56 & 0.001 & -410.5 & 0.98 & -8.22 & 0.002 & -627.1 &
    1.49 & 0.0&\cite{BiOX}\\
    BiOI &\cite{BiOX}& -10.69 & 0.001 & -372.5 & 0.89 & -7.00 & 0.002 & -569.4 &
    1.36 & 0.0&\cite{BiOX}\\\hline\hline
  \end{tabular}
\end{table}

For all As and Sb compounds, we observe good agreement with experimental data. The differences are between 14 and 20\% for the NQMs, which is in agreement with the error for DFT calculations on EFGs in the literature~\cite{petrilli1998electric}. The error also includes temperature and other effects. 
For $\beta$-$^{121}$SbBr$_3$, experimental data is only available at 300~K instead of at 77~K. 
If we assume a similar temperature behavior as $\alpha$-$^{121}$SbBr$_3$, the ratio between theoretical and experimental NQMs should increase to about 1.15. The agreement for $\eta$ is generally good, although this quantity potentially benefits more from error cancellation. The only exception is $\alpha$-$^{121}$SbBr$_3$, with a measured value of 0.08 compared to the calculated value of 0.155.

Turning now to the bismuth compounds, the results for BiCl$_3$ are in good agreement with experiment. An overshoot of 17\% is within expectations and $\eta$ differs by only $0.02$. For BiBr$_3$, neither the calculated EFG nor the asymmetry parameter are in agreement with experiment. However, if we used a different assignment of the experimental data, as discussed below, the differences disappear and the agreement between theory and experiment is comparable to that of BiCl$_3$.
Both BiCl$_3$ and BiBr$_3$, assuming the different frequency assignment, support a value of -420 mb~\cite{barzakh2021large,dognon2023determining} for the NQM of $^{209}$Bi.

The results for the bismuth oxyhalides are worse. The calculated EFGs decrease for the heavier halides, following the same trend as observed for the other systems. However, when combined with the experimental data, the resulting NQMs for $^{209}$Bi are erratic. BiOF yields an expected error of 10\%, but the other three systems are much worse, without any discernible trend. Cl and Br have an overshoot of almost 50\%, while I is only off by 36\%.
These are beyond expected variation both in relative accuracy as well as variation between the different oxyhalides, and we are currently unable to resolve these discrepancies.
Whereas BiOF also supports a value of -420 mb  for the NQM, it's reliability is questionable given the differences observed for the other three BiOXs.

As seen from Table~\ref{tab:results}, the agreement between the calculated and experimental values for $\eta$ is good
for all compounds, except for BiBr$_3$, which shows a large disagreement. This general agreement may be due to error cancellation considering that $\eta$ involves a ratio between different EFG components.
Furthermore, across all other compounds the EFG decreases for heavier halides. In contrast, the experimental data for BiBr$_3$ suggest a small increase in the EFG compared to BiCl$_3$.

To investigate this conundrum, we have reoptimized the values for the nuclear quadrupole coupling constant and $\eta$ from the experimentally measured frequencies. For most systems, we obtain a better agreement with the measured frequencies with a pure quadrupole spectra from this refitting process.
While the actual values of the EFGs do not change significantly, the refitting leads to a reduction in the experimental uncertainties. For example, for BiCl$_3$, the agreement is below 0.1\%. The only major change is for $^{121}$Sb in the $\alpha$-structure, where the value changes from 343.95 MHz to 331.41 MHz, which appears to be due to a mistake in one of the tables of Ref.~\cite{SbBrdata}.

The agreement is also excellent for BiBr$_3$, however, only three out of the four frequencies were observed. Therefore the frequency assignment is ambiguous, and different choices are possible. We find a different assignment for $\eta=0.840$.
This value of $\eta$ agrees with our calculated value for $\eta$, gives a good agreement for the NQM of $^{209}$Bi with the same overshoot as seen for the arsenic and antimony compounds, and follows the trend of lowering the EFG for heavier halides.
However, this fit has a significantly higher $\chi^2$ value of $4\cdot 10^{-1}$ compared to the experimental fit of $6\cdot 10^{-4}$. For the other compounds, we obtain $\chi^2$ values ranging from $2\cdot 10^{-2}$ to $4\cdot 10^{-6}$, making this fit an outlier. We are at this point unable to provide an decisive conclusion on the origin of this discrepancy.

\begin{table}[]
  \caption{Results for EFGs [$10^{21}$ V/$m^2$] and asymmetry parameters $\eta$ in gas phase, for isolated molecular structures using solid-state geometries as well as for solids. For details, see text.}
  \label{tab:mol}
\begin{tabular}{lcccccc}
\hline
\hline
Compound & \multicolumn{3}{c}{EFG} & \multicolumn{3}{c}{$\eta$} \\
         &gas& 'molecule'     & solid    &gas & 'molecule'             & solid            \\ \hline
AsCl$_3$            &-16.95&-19.22&-18.47&0&0.012&0.027 \\
AsBr$_3$            &-13.86&-14.79&-13.65&0&0.048&0.082 \\
SbCl$_3$            &-25.15&-28.71&-25.60&0&0.090&0.170 \\
$\alpha$-SbBr$_3$   &-21.36&-24.99&-20.76&0&0.10&0.155 \\
$\beta$-SbBr$_3$    &\footnote{Only one configuration in gas phase.}&-24.61&-21.50& &0.11&0.202 \\
BiCl$_3$            &-30.95&-39.77&-27.36&0&0.229&0.565 \\
BiBr$_3$            &-25.55&-32.67&-21.81&0&0.171&0.850 \\ 
\hline\hline
\end{tabular}
\end{table}

To estimate a value for the NQM of $^{209}$Bi based on the
BiCl$_3$ data, we apply a scaling factor to the calculated value. The
scaling factor is taken as the  the average ratios of
calculated and experimental NQMs for AsCl$_3$, AsBr$_3$, $^{121}$SbCl$_3$, and
the two sites in $^{123}$Sb$_4$O$_5$Cl$_2$, as the temperature is the same
for all these studies, and good agreement between calculated and
measured $\eta$ is observed. We then obtain a NQM of -428(17) mbarn,
with the number in parenthesis representing the standard deviation,
heavily influenced by the result for AsBr$_3$.

We have also implemented the calculation of EFGs in the molecular
ReSpect program. As we have a uniform level of
theory for molecules and solids, and many of the pnicogen halides are
molecular solids, we have also explored the effects of crystal packing
as well as the geometry effects arising from differences in bond
lengths and bond angles going from the gas-phase structure to the
structure observed in the solid state. We have thus performed
calculations on individual molecules by taking the positions of one pnicogen and its three nearest bonding halides from the respective solid-state structures to study  solid-state effects, as well as
optimized the molecular structure in the gas phase to explore the
relaxation effects. All computational details are the same, including XC
functional and basis sets.
The calculated EFGs and asymmetry parameters are collected
in Table~\ref{tab:mol}.\\
There are significant effects both from changing the bond length going from gas phase to the isolated molecule retaining the solid-state structure,  as well as due to neighboring atoms in the solid. Only the solid state calculations give results for both EFGs and $\eta$ that are in agreement with experiment, illustrating the need to use both a proper geometry as well as include the solid-state effects.
The increasing difference in EFGs from isolated solid-state structure molecules to the solid phase for the heavier compounds is not directly caused by relativity. This is rather due to the closer crystal packing, leading to larger crystal-packing effects. Both the pnicogen-pnicogen distances as well as the distance to the nearest non-bonding halides are for these heavier congeners smaller.
$\eta$ in the gas phase is zero due to the trigonal pyramidal geometry,~\cite{pyykko2008deuteron} whereas the value of
 $\eta$ for isolated solid-state structures and the solids show that it depends strongly on the environment and is not only a results of the distorted trigonal pyramid in solids.

\section{Concluding remarks}
\label{sec:conclusions}

We have presented the theory for calculating EFGs in solids within the all-electron, four-component framework using GTOs as implemented in the \textsc{ReSpect} program.
For As and Sb, these values are well established and we can  use them to determine the accuracy of EFGs calculated using our approach, and in particular the inherent error of the DFT approximation applied in this work. We have obtained good agreement for various systems, with errors in the range of 14-20\%, in line with other findings in the literature.

We have demonstrated that our solid-state approach for calculating electric field gradients is reliable enough to determine
the nuclear quadrupole moment from observed nuclear quadrupole coupling constants.
For Q(Bi), our current, independent, result is -428(17) mb.
It is comparable to the published 'World average' of -420(17) mb of Barzakh {\it et al.\/}~\cite{barzakh2021large}
or the 'molecular' values of Teodoro and Haiduke (-420(8) mb)~\cite{teodoro2013nuclear} or Dognon and Pyykk\"{o} (-422(3) mb)~\cite{dognon2023determining} and others, as well as with recent nuclear calculations by Karayonchev {\it et al.\/} (-428 mb)~\cite{karayonchev2019lifetimes}.
Note that the old atomic value of  Bieron and Pyykkö (-516(15) mb)~\cite{bieron2001nuclear} suffered from
a too limited electron correlation space. Further values and a new World average Q(Bi) should be proposed later.


However, the nuclear quadrupole moments derived from bismuth oxyhalides behave erratically, even though the calculated EFGs appear reasonable. Although the results for BiOF also supports the molecular value for the NQM of $^{209}$Bi,  its reliability is in question until the disagreements for the other oxyhalides are resolved. We recommed that new experimental studies are performed for bismuth oxyhalides.

\section{Acknowledgment}

This work was supported by the Research Council of Norway through its Centres of Excellence scheme, project number 262695, its FRIPRO grant nr. 315822, and the Mobility Grant scheme (project no. 301864). For the computational and data storage resources, we acknowledge the support of  Sigma2---the National Infrastructure for High Performance Computing and Data Storage in Norway, Grant Nos.~NN4654K and~NN14654K.
In addition, MR acknowledges funding from the European Union’s Horizon 2020 research and innovation program under the Marie Skłodowska-Curie Grant Agreement No. 945478 (SASPRO2), and the Slovak Research and Development Agency (Grant No. APVV-21-0497). PP is Professor Emeritus at University of Helsinki.

\end{document}